\documentclass[12pt,thmsa]{article}

\input{tcilatex}
\begin{document}

\setcounter{page}{0} \topmargin 0pt \oddsidemargin 5mm \renewcommand{%
\thefootnote}{\fnsymbol{footnote}} \newpage \setcounter{page}{0} 
\begin{titlepage}
\begin{flushright}
Berlin Sfb288 Preprint  \\
hep-th/0001128\\
\end{flushright}
\vspace{0.5cm}
\begin{center}
{\Large {\bf Colour valued Scattering Matrices} }

\vspace{0.8cm}
{\large A. Fring and C. Korff }

\vspace{0.5cm}
{\em Institut f\"ur Theoretische Physik,
Freie Universit\"at Berlin\\ 
Arnimallee 14, D-14195 Berlin, Germany }
\end{center}
\vspace{0.2cm}
 
\renewcommand{\thefootnote}{\arabic{footnote}}
\setcounter{footnote}{0}

\begin{abstract}
We describe a general construction principle which allows to add colour values to
a coupling constant dependent scattering matrix. As a concrete realization of this
mechanism we provide a new type of S-matrix which generalizes the one of affine Toda 
field theory, being related to a pair of Lie algebras. A characteristic feature of this 
S-matrix is that in general it violates  parity invariance. For particular choices of the two
Lie algebras involved this scattering
matrix coincides with the one related to the scaling models described by  the minimal 
affine Toda S-matrices and for other choices  with the one of the Homogeneous
sine-Gordon models with vanishing resonance parameters. We carry out the 
thermodynamic Bethe ansatz and identify the corresponding ultraviolet effective central charges.
\medskip
\par\noindent
PACS numbers: 11.10Kk, 11.55.Ds
\end{abstract}
\vfill{ \hspace*{-9mm}
\begin{tabular}{l}
\rule{6 cm}{0.05 mm}\\
Fring@physik.fu-berlin.de \\
Korff@physik.fu-berlin.de \\
\end{tabular}}
\end{titlepage}
\newpage

\section{Introduction}

The bootstrap principle \cite{boot} has turned out to be a successful method
to compute scattering matrices in 1+1-dimensions. Solving the set of
bootstrap equations and giving a consistent explanation to the singularity
structure in the complex rapidity plane, the scattering matrices are
determined uniquely up to the so-called CDD-factors \cite{CDD}. The latter
factors are constituted in such a way that they solve the bootstrap
equations but do not introduce additional poles in the physical sheet.
Therefore they are neglected in most situations. However, they may also by
utilized in order to include coupling constants into the scattering
matrices, as for instance in \cite{TodaS}. We will show in the following
that the CDD-factors can also be employed consistently to add colour values
to the scattering matrices. In the context of the Homogeneous sine-Gordon
models Fern\'{a}ndez-Pousa and Miramontes \cite{HSGS} proposed a new type of
S-matrix which violates parity invariance and describes the scattering of
particles which carry two quantum numbers. The main quantum number governs
the fusing structure while for certain values of the colour quantum numbers
the particles interact solely via a CDD-factor, which could be trivial in
some cases. We will provide a systematic construction principle for colour
valued scattering matrices and give explicit realizations which include the
ones of \cite{HSGS} as a particular case. These type of theories are related
to two different Lie algebras \textbf{g }and \textbf{\~{g}}, where the
former relates to the main and the latter to the colour quantum number. We
refer to these theories by \textbf{g\TEXTsymbol{\vert}\~{g}.}

Our manuscript is organized as follows: In section 2 we describe the general
construction principle which attaches colour values to an S-matrix and
provide a concrete realization of this. In section 3 we carry out a
TBA-analysis in order to identify the corresponding ultraviolet effective
central charges. We provide an explicit example in section 4. In section 5
we state our conclusions.

\section{Construction Principle}

We recall that the two-particle S-matrix which describes the scattering
between particles of type $a$ and $b$ as a function of the rapidity
difference $\theta $ is often\footnote{%
This is not necessarily the case as for instance for affine Toda field
theories related to non-simply laced Lie algebras, which was first worked
out in \cite{Gust}. See also \cite{FKS2} and references therein for a recent
treatment of these type of theories.} of the general form $S_{ab}(\theta
)=S_{ab}^{\min }(\theta )S_{ab}^{\text{CDD}}(\theta ,B)$. Here $S_{ab}^{\min
}(\theta )$ is the so-called minimal S-matrix, related for instance to
scaling theories of statistical models \cite{Zper}, which satisfies the
unitarity, crossing and fusing bootstrap equations 
\begin{equation}
S_{ab}(\theta )S_{ba}(-\theta )=1,\qquad S_{\bar{a}b}(\theta )=S_{ba}(i\pi
-\theta ),\qquad \prod_{l=a,b,c}S_{dl}(\theta +\eta _{l})=1\,\,\,,
\label{boot}
\end{equation}
with $\eta _{l}$ being the fusing angles. The CDD-factor $S_{ab}^{\text{CDD}%
}(\theta ,B)$ depends on the effective coupling constant and is chosen in
such a way that it also satisfies these equations without introducing
additional poles in the physical sheet, i.e. $0\leq \func{Im}\theta \leq \pi 
$. We may now modify the usual expression to 
\begin{equation}
\hat{S}_{ab}^{ij}(\theta )=S_{ab}^{\min }(\theta )S_{ab}^{\text{CDD}}(\theta
,B_{ij})\,\,.  \label{S1}
\end{equation}
Here we have introduced an additional dependence of the effective coupling
constant on the quantum numbers $i$ and $j$, which we refer to as colours.
It is clear by construction that $\hat{S}_{ab}^{ij}(\theta )$ also satisfies
the crossing, unitarity and fusing bootstrap equations, but now each
particle carries two quantum numbers $(a,i)$, which may take their values in
different ranges, for definiteness say $1\leq a\leq \ell $ and $1\leq i\leq 
\tilde{\ell}$. This means, now we have in total $\tilde{\ell}\times \ell $
different particle types. Alternatively, we can define an S-matrix which
coincides with one or the other factor in (\ref{S1}) for certain colour
values 
\begin{equation}
S_{ab}^{ij}(\theta )=\QATOPD\{ . {S_{ab}^{\min }(\theta
)=(S_{ab}^{CDD}(\theta ,B_{ii}=0))^{-1}\qquad \qquad \text{for }%
i=j}{S_{ab}^{CDD}(\theta ,B_{ij})\qquad \qquad \qquad \,\qquad \qquad \quad
\quad \text{for }i\neq j}\,\,\,.  \label{S}
\end{equation}
This means whenever $i=j$ we simply have $\tilde{\ell}$ copies of theories
which interact via a minimal scattering matrix and for $i\neq j$ the
particles interact purely via a CDD-factor. Clearly by construction also (%
\ref{S}) satisfies the consistency equations (\ref{boot}). It should be
noted here that (\ref{S1}) and (\ref{S}) still describe scattering processes
for which backscattering is absent. Hence, these type of colour values play
a different role as those which occur for instance in S-matrices related to
affine Toda field theories \cite{ATFT} with purely imaginary coupling
constant, e.g. \cite{nondiag}. Despite the fact that the relative mass
spectra related to (\ref{S}) are degenerate, this is consistent when we
encounter $\tilde{\ell}$ different overall mass scales or the particles have
different charges.

We will now generalize the structure just outlined, which we already
encountered in \cite{CFKM}, and provide a concrete realization for $%
S_{ab}^{ij}(\theta )$, which is of affine Toda field theory type, involving
a pair of simply laced Lie algebras. It is clear, however, from our previous
comments that the forms (\ref{S1}) and (\ref{S}) are of a more general
nature. We associate the main quantum numbers $a,b$ to the vertices of the
Dynkin diagram of a simply laced Lie algebra \textbf{g }of rank $\ell $ and
the colour quantum numbers $i,j$ to the vertices of the Dynkin diagram of a
simply laced Lie algebra \textbf{\~{g}} of rank $\tilde{\ell}$ and refer
from now on to these theories as \textbf{g\TEXTsymbol{\vert}\~{g}}\footnote{%
This should of course not be understood as a coset.}.

We define now the general building blocks 
\begin{equation}
\left[ x,B\right] _{\theta ,ij}=e^{\frac{i\pi x\varepsilon _{ij}}{h}}\left( 
\frac{\sinh \tfrac{1}{2}(\theta +i\pi (x-1+B)/h)\sinh \tfrac{1}{2}(\theta
+i\pi (x+1-B)/h)}{\sinh \tfrac{1}{2}(\theta -i\pi (x-1+B)/h)\sinh \tfrac{1}{2%
}(\theta -i\pi (x+1-B)/h)}\right) ^{\frac{1}{2}}\,\,.  \label{bl}
\end{equation}
with $\varepsilon _{ij}$ being the anti-symmetric tensor, i.e. $\varepsilon
_{ij}=-\varepsilon _{ji}$. It is this property of the $\varepsilon $'s which
is responsible for the parity breaking of the S-matrix. This block has the
obvious properties 
\begin{equation}
\left[ x,B\right] _{\theta ,ij}\left[ x,B\right] _{-\theta ,ji}=1\qquad 
\text{and\qquad }\left[ h-x,B=1\right] _{\theta ,ij}=\left[ x,B=1\right]
_{i\pi -\theta ,ji}\,\,.  \label{prop}
\end{equation}
We understand here in a slightly loose notation that in the second equality
we first take the square root and thereafter perform the shifts in the
arguments. Note further that the order of the colour values is relevant.
From (\ref{bl}) we construct the \textbf{g\TEXTsymbol{\vert}\~{g}}%
--scattering matrix 
\begin{equation}
S_{ab}^{ij}(\theta )=\prod\limits_{q=1}^{h}\left[ 2q-(c_{a}+c_{b})/2),\tilde{%
I}_{ij}\right] _{\theta ,ij}^{-\frac{\tilde{K}_{ij}}{2}\lambda _{a}\cdot
\sigma ^{q}\gamma _{b}}\,\,.  \label{Snew}
\end{equation}
This is of the form (\ref{S}) apart from a phase factor and a square root
taken when $i\neq j$. Here the $\lambda _{a}$'s are fundamental weights, the 
$\gamma _{a}$'s are simple roots times a colour value $c_{a}=\pm 1$, $h$ is
the Coxeter number and $\sigma $ is the Coxeter element related to the Lie
algebra \textbf{g}. $\tilde{K}$ is the Cartan matrix and $\tilde{I}=2-\tilde{%
K}$ the incidence matrix of the \textbf{\~{g}} related Dynkin diagram. For
more details on the notation and properties of the quantities involved see 
\cite{FO,FKS2}. For $i=j$ we recover with $\tilde{I}_{ii}=0$ and $\tilde{K}%
_{ii}=2$ the known form of the minimal scattering matrix of affine Toda
field theory. Whenever $i$ and $j$ are not linked on the \textbf{\~{g}}%
--Dynkin diagram $S$ becomes 1, i.e. the particles interact freely. Instead
when $i$ and $j$ are linked on the \textbf{\~{g}--}Dynkin diagram, we have $%
\tilde{I}_{ij}=1$ and $\tilde{K}_{ij}=-1$ such that the corresponding blocks
are inverted. Comparing (\ref{bl}) with the conventional minimal blocks, we
have introduced the parity breaking phase factor and also taken the square
root to minimize the powers of the poles since in $\left[ x,B=1\right]
_{\theta ,ij}$ the two factors in the denominator and as well as in the
numerator coincide. Hence for $i\neq j$ the expression (\ref{Snew})
corresponds to the square root of the usual affine Toda field theory
CDD-factor for $B=1$. It is this operation of taking the square root which
is the reason for the occurrence of the phase factor in (\ref{bl}), since
only with its presence the consistency equations are satisfied.

There is no need to introduce the phase to satisfy the unitarity equation in
(\ref{boot}), since the first property in (\ref{prop}) is satisfied with or
without it. However, already in order to satisfy the crossing relation the
introduction of the phase factor is crucial since the second property in (%
\ref{prop}), which is needed to establish it, only holds when it is
included. Assuming the validity of the ADE-fusing rules one may verify by
the usual shifting arguments, e.g. \cite{FO,FKS2}, that the fusing bootstrap
equations are satisfied. It is further clear that (\ref{Snew}) is hermitian
analytic \cite{HERMAN}.

For many applications, like the thermodynamic Bethe ansatz or form factors,
it is most convenient to employ the scattering matrix in form of an integral
representation instead of the blockform (\ref{bl}). In \cite{FKS1,FKS2} it
was demonstrated how to derive one formulation from the other and by
specifying the analysis in there to the present situation it follows
immediately that we can express the scattering matrix (\ref{S})
alternatively as

\begin{equation}
S_{ab}^{ij}(\theta )=e^{i\pi \varepsilon _{ij}K_{ab}^{-1}}\exp
\int\limits_{-\infty }^{\infty }\frac{dt}{t}\,\left( 2\cosh \frac{\pi t}{h}-%
\tilde{I}\right) _{ij}\left( 2\cosh \frac{\pi t}{h}-I\right)
_{ab}^{-1}\,\,e^{-it\theta }\,\,\,.  \label{Sint}
\end{equation}

The pre-factor results from a similar computation as may be found in section
4.2.1. of \cite{FKS1}.

We note that when we choose \textbf{\~{g}} to be $A_{1}$ the colour values
become identical for all particles and the system reduces to the one
described by $S_{ab}^{\min }(\theta )$. This is the only example for which (%
\ref{Snew}), (\ref{Sint}) does not violate the parity invariance. Choosing
instead \textbf{g }to be $A_{n}$ we recover the S-matrix of the Homogeneous
sine-Gordon models for vanishing resonance parameter at level $(n+1)$ \cite
{HSGS,CFKM}.

Similar as in the case for which the universal scattering matrix (\ref{Snew}%
) coincides with models already known, also all S-matrix elements which
belong to the new theories are well-behaved meromorphic functions. At first
sight the power $1/2$ in the definition of the building block (\ref{bl})
seems to suggest the presence of square root branch cuts. For the \textbf{g%
\TEXTsymbol{\vert}}$A_{1}$--model the $1/2$ is familiar for instance from 
\cite{FO} where it is kept as a power in relation (\ref{Snew}). A detailed
analysis which explains how the building blocks combine to meromorphic
functions may be found in there. For the case $B=\tilde{I}_{ij}=1$ the
square root can be taken directly in (\ref{bl}) and the remaining power $1/2$
in (\ref{Snew}) is once again compensated by the same mechanism as in \cite
{FO}.

It is straightforward to include also resonance parameters into the
scattering matrix (\ref{Snew}), (\ref{Sint}) which could in principle be
colour value dependent and may also break the parity invariance \cite
{HSGS,CFKM}.

\section{TBA Analysis for the g\TEXTsymbol{\vert}\~{g} S-matrix}

According to the standard arguments of the thermodynamic Bethe ansatz \cite
{TBAZam1} the TBA-equations for a system which interacts dynamically via the
scattering matrix (\ref{Sint}) and statistically via Fermi statistics read 
\begin{equation}
rm_{a}^{i}\cosh \theta =\varepsilon _{a}^{i}(\theta
)+\sum\limits_{b=1}^{\ell }\sum\limits_{j=1}^{\tilde{\ell}%
}\int\limits_{-\infty }^{\infty }d\theta ^{\prime }\varphi _{ab}^{ij}(\theta
-\theta ^{\prime })\ln \left( 1+e^{-\varepsilon _{b}^{j}(\theta ^{\prime
})}\right) \,\,.  \label{TBA}
\end{equation}
Here $r$ is the inverse temperature and $m_{a}^{i}$ the mass of particle $%
(a,i)$. The pseudoenergies are denoted as usual by $\varepsilon
_{a}^{i}(\theta )$ and the kernels are obtained from (\ref{Sint}) 
\begin{equation}
\varphi _{ab}^{ij}(\theta )=-i\frac{d}{d\theta }\ln S_{ab}^{ij}(\theta
)=\int\limits_{-\infty }^{\infty }dt\left[ \delta _{ab}\delta _{ij}-\left(
2\cosh \frac{\pi t}{h}-\tilde{I}\right) _{ij}\left( 2\cosh \frac{\pi t}{h}%
-I\right) _{ab}^{-1}\,\right] \,\,e^{-it\theta }\,\,.  \label{9}
\end{equation}
One of the most direct informations the thermodynamic Bethe ansatz provides
is the effective central charge $c_{\text{eff}}=c-24h_{0}$ of the underlying
ultraviolet conformal field theory, with $c$ being the Virasoro central
charge and $h_{0}$ the smallest conformal dimension of the theory. Then,
provided that the solutions of the TBA-equation develop the usual ``plateau
behaviour''\footnote{%
This is not always the case as for instance in affine Toda field theories
with generic effective coupling constant \cite{FKS1}.}, e.g. \cite{TBAZam1},
one may approximate $\varepsilon _{a}^{i}(\theta )=\varepsilon _{a}^{i}=$ $%
const$ in a large region for $\theta $ when $r$ is small. By standard TBA
arguments \cite{TBAZam1} follows that the effective central charge is
expressible as 
\begin{equation}
c_{\text{eff}}=\frac{6}{\pi ^{2}}\sum\limits_{a=1}^{\ell }\sum\limits_{i=1}^{%
\tilde{\ell}}\mathcal{L}\left( \frac{x_{a}^{i}}{1+x_{a}^{i}}\right)
\label{ceff}
\end{equation}
with $\mathcal{L}(x)=\sum_{n=1}^{\infty }x^{n}/n^{2}+\ln x\ln (1-x)/2$
denoting Rogers dilogarithm \cite{Log} where the $x_{a}^{i}=\exp
(-\varepsilon _{a}^{i})$ are obtained as solutions from the constant
TBA-equations in the form 
\begin{equation}
x_{a}^{i}=\prod\limits_{b=1}^{\ell }\prod\limits_{j=1}^{\tilde{\ell}%
}(1+x_{b}^{j})^{N_{ab}^{ij}}\,\,.  \label{cTBA}
\end{equation}
The matrix $N_{ab}^{ij}$ is defined via the asymptotic behaviour of the
scattering matrix which for the case at hand may be read off directly from (%
\ref{9}) 
\begin{equation}
N_{ab}^{ij}=\frac{1}{2\pi }\int\limits_{-\infty }^{\infty }d\theta
\,\,\varphi _{ab}^{ij}(\theta )=\,\delta _{ab}\delta _{ij}-K_{ab}^{-1}\tilde{%
K}_{ij}\,.
\end{equation}
In regard to finding explicit solutions for the set of coupled equations (%
\ref{cTBA}), it turns out to be convenient to introduce new variables
because they may be related to Weyl characters of the Lie algebra \textbf{g }%
or \textbf{\~{g}}. Following \cite{Resh,Kun} we define 
\begin{equation}
Q_{a}^{i}=\prod_{b=1}^{\ell }(1+x_{b}^{i})^{K_{ab}^{-1}}\qquad
\Leftrightarrow \qquad x_{a}^{i}=\prod_{b=1}^{\ell }\left( Q_{b}^{i}\right)
^{K_{ab}}-1
\end{equation}
such that the constant TBA-equations (\ref{cTBA}) acquire the more symmetric
form 
\begin{equation}
\prod_{b=1}^{\ell }\left( Q_{b}^{i}\right) ^{I_{ab}}+\prod_{j=1}^{\tilde{\ell%
}}\left( Q_{a}^{j}\right) ^{\tilde{I}_{ij}}=\left( Q_{a}^{i}\right) ^{2}\,\,.
\label{QQ}
\end{equation}
The effective central charge (\ref{ceff}) is then expressible in various
equivalent ways 
\begin{eqnarray}
c_{\text{eff}}^{\mathbf{g}|\mathbf{\tilde{g}}} &=&\frac{6}{\pi ^{2}}%
\sum\limits_{a=1}^{\ell }\sum\limits_{i=1}^{\tilde{\ell}}\mathcal{L}\left(
1-\prod_{b=1}^{\ell }\left( Q_{b}^{i}\right) ^{-K_{ab}}\right) =\medskip
\ell \tilde{\ell}-\frac{6}{\pi ^{2}}\sum\limits_{a=1}^{\ell
}\sum\limits_{i=1}^{\tilde{\ell}}\mathcal{L}\left( 1-\prod_{j=1}^{\tilde{\ell%
}}\left( Q_{a}^{j}\right) ^{-\tilde{K}_{ij}}\right) \,\,  \label{cc} \\
&=&\ell \tilde{\ell}-\frac{6}{\pi ^{2}}\sum\limits_{a=1}^{\ell
}\sum\limits_{i=1}^{\tilde{\ell}}\mathcal{L}\left( \prod_{b=1}^{\ell }\left(
Q_{b}^{i}\right) ^{-K_{ab}}\right) =\frac{6}{\pi ^{2}}\sum\limits_{a=1}^{%
\ell }\sum\limits_{i=1}^{\tilde{\ell}}\mathcal{L}\left( \prod_{j=1}^{\tilde{%
\ell}}\left( Q_{a}^{j}\right) ^{-\tilde{K}_{ij}}\right)  \label{cc2}
\end{eqnarray}
where we used the well-known identity $\mathcal{L}(x)+\mathcal{L}(1-x)=\pi
^{2}/6$, see e.g. \cite{Log}. It is also clear that having solved the
equations (\ref{QQ}) for the case \textbf{g}\TEXTsymbol{\vert}\textbf{\~{g} }%
we have immediately a solution for the case \textbf{\~{g}\TEXTsymbol{\vert}g 
}simply by interchanging the role of the two algebras. Supposing now that $%
c_{\text{eff}}^{\mathbf{g}|\mathbf{\tilde{g}}}=\mu \,c_{\text{eff}}^{\mathbf{%
\tilde{g}}|\mathbf{g}}$ for some unknown constant $\mu $, it follows
directly from (\ref{cc}) and (\ref{cc2}) that $c_{\text{eff}}^{\mathbf{g}|%
\mathbf{\tilde{g}}}=\mu \ell \tilde{\ell}/(1+\mu )$. We conjecture now this
constant to be $\mu =\tilde{h}/h$ such that 
\begin{equation}
c_{\text{eff}}^{\mathbf{g|\tilde{g}}}=\frac{\ell \tilde{\ell}\,\tilde{h}}{h+%
\tilde{h}}\,\,\,.  \label{ceffn}
\end{equation}
As expected from the observations concerning the scattering matrix we
recover several known cases when we fix some of the algebras. For instance
we obtain $c_{\text{eff}}^{\mathbf{g|}A_{1}}=2\ell /(h+2)$ which is the well
known formula for the effective central charge of the minimal affine Toda
theories. Furthermore we recover the effective central charge for the
Homogeneous sine-Gordon models $c_{\text{eff}}^{A_{n}|\mathbf{\tilde{g}}}=n%
\tilde{\ell}\tilde{h}/(n+1+\tilde{h})$ \cite{CFKM}. It should be noted that
this is independent of whether a resonance parameter is present or not
despite the fact that the TBA-equation are not parity invariant in that
case, see \cite{CFKM}. Numerically we also solved (\ref{QQ}) explicitly for
numerous examples with \textbf{g}$\neq \!\!A_{n}$ and confirmed (\ref{ceffn}%
).

\begin{center}
\begin{tabular}{|c||c|c|c|c|c|}
\hline
\textbf{g\TEXTsymbol{\vert}\~{g}} & $A_{m}$ & $D_{m}$ & $E_{6}$ & $E_{7}$ & $%
E_{8}$ \\ \hline\hline
$A_{n}$ & $\frac{nm\,(m+1)}{n+m+2}$ & $\frac{nm\,(2m-2)}{n+2m-1}$ & $\frac{%
72\,n}{n+13}$ & $\frac{126\,n}{n+19}$ & $\frac{240\,n}{n+31}$ \\ \hline
$D_{n}$ & $\frac{nm\,(m+1)}{2n+m-1}$ & $\frac{nm\,(m-1)}{n+m-2}$ & $\frac{%
36\,n\,}{n+5}$ & $\frac{63\,n\,}{n+8}$ & $\frac{120\,\,n\,}{n+14}$ \\ \hline
$E_{6}$ & $\frac{6\,m\,(m+1)}{m+13}$ & $\frac{6\,m\,(m-1)}{m+5}$ & $18$ & $%
\frac{126}{5}$ & $\frac{240}{7}$ \\ \hline
$E_{7}$ & $\frac{7\,m\,(m+1)}{m+19}$ & $\frac{7\,m\,(m-1)}{m+8}$ & $\frac{84%
}{5}$ & $\frac{49}{2}$ & $35$ \\ \hline
$E_{8}$ & $\frac{8\,m\,(m+1)}{m+31}$ & $\frac{8\,m\,(m-1)}{m+14}$ & $\frac{96%
}{7}$ & $21$ & $32$ \\ \hline
\end{tabular}

\vspace{0.4cm} {\small Table1: Effective central charges }$c_{\text{eff}}^{%
\mathbf{g|\tilde{g}}}$ {\small of the}\textbf{\ g\TEXTsymbol{\vert}\~{g}}%
{\small -theories.}
\end{center}

\section{An Explicit example: $D_{4}|D_{4}$}

In order to illustrate the working of our general formulae it is instructive
to evaluate them for a concrete model. We chose the $D_{4}|D_{4}$-model
which is an example for a therory hitherto unknown. The model contains 16
different particles labeled by $(a,i)$ with $1\leq a,i\leq 4$. The Coxeter
number is $6$ for $D_{4}$. Naming the central particle in the $D_{4}$%
--Dynkin diagram by 2 the S-matrix elements according to (\ref{Snew}) are
computed to 
\[
\begin{tabular}{ll}
$S_{11}^{ii}(\theta )=S_{33}^{ii}(\theta )=S_{44}^{ii}(\theta
)=[1,0]_{\theta ,ii}^{2}[5,0]_{\theta ,ii}^{2}$ & $\text{for }%
i=1,2,3,4,\smallskip $ \\ 
$S_{12}^{ii}(\theta )=S_{23}^{ii}(\theta )=S_{24}^{ii}(\theta
)=[2,0]_{\theta ,ii}^{2}[4,0]_{\theta ,ii}^{2}$ & $\text{for }%
i=1,2,3,4,\smallskip $ \\ 
$S_{13}^{ii}(\theta )=S_{14}^{ii}(\theta )=S_{34}^{ii}(\theta
)=[3,0]_{\theta ,ii}^{2}$ & $\text{for }i=1,2,3,4,\smallskip $ \\ 
$S_{22}^{ii}(\theta )=[1,0]_{\theta ,ii}^{2}[3,0]_{\theta
,ii}^{4}[5,0]_{\theta ,ii}^{2}$ & $\text{for }i=1,2,3,4,\smallskip $ \\ 
$S_{11}^{2j}(\theta )=S_{33}^{2j}(\theta )=S_{44}^{2j}(\theta
)=[1,1]_{\theta ,2j}^{-2}[5,1]_{\theta ,2j}^{-2}$ & $\text{for }%
j=1,3,4,\smallskip $ \\ 
$S_{12}^{2j}(\theta )=S_{23}^{2j}(\theta )=S_{24}^{2j}(\theta
)=[2,1]_{\theta ,2j}^{-2}[4,1]_{\theta ,2j}^{-2}\qquad $ & $\text{for }%
j=1,3,4,\smallskip $ \\ 
$S_{13}^{2j}(\theta )=S_{14}^{2j}(\theta )=S_{34}^{2j}(\theta
)=[3,1]_{\theta ,2j}^{-2}$ & $\text{for }j=1,3,4,\smallskip $ \\ 
$S_{22}^{2j}(\theta )=[1,1]_{\theta ,2j}^{2}[3,1]_{\theta
,2j}^{4}[5,1]_{\theta ,2j}^{2}$ & $\text{for }j=1,3,4,\smallskip $ \\ 
$S_{ab}^{ij}(\theta )=1$ & $\text{for }a,b=1,2,3,4;\,\,i\neq j;\,i,j\neq 2\,.
$%
\end{tabular}
\]
The solutions to the constant TBA-equations (\ref{cTBA}) read 
\begin{eqnarray}
x_{1}^{1}
&=&x_{3}^{1}=x_{4}^{1}=x_{1}^{3}=x_{1}^{4}=x_{3}^{3}=x_{3}^{4}=x_{3}^{4}=x_{4}^{4}=x_{2}^{2}=1
\\
x_{2}^{1} &=&x_{2}^{3}=x_{2}^{4}=1/2 \\
x_{1}^{2} &=&x_{3}^{2}=x_{4}^{2}=2
\end{eqnarray}
such that the effective central charge according to (\ref{ceff}) is 
\begin{equation}
c_{\text{eff}}=\frac{6}{\pi ^{2}}\left( 10\mathcal{L}\left( \frac{1}{2}%
\right) +3\mathcal{L}\left( \frac{2}{3}\right) +3\mathcal{L}\left( \frac{1}{3%
}\right) \right) =8\,\,.
\end{equation}
This result confirms the general formula (\ref{ceffn}).

\section{Conclusions}

We have shown that the proposed scattering matrices (\ref{Snew}), (\ref{Sint}%
) provide consistent solutions of the bootstrap equations (\ref{boot}). In
comparison with (\ref{S}) we have taken the square root of the CDD-factor
which lead to the introduction of the non-trivial parity breaking phase
factors. The main motivation for this was to recover the known scattering
matrices which were mentioned at the end of section 2. It is clear though
that when we view (\ref{S1}) as the usual affine Toda field theory
scattering matrix related to simply laced algebras we can write down
immediately colour valued S-matrices related to two different algebras. When
we do not take the square root this is straightforward and also works for
theories related to non-simply laced algebras. We leave a systematic
investigation of these type of theories for future investigations.

A further open question is to identify the corresponding Lagrangian for the 
\textbf{g\TEXTsymbol{\vert}\~{g}}-theories. The knowledge of the ultraviolet
central charge (\ref{ceffn}) will certainly be useful in this search since
it provides the renormalization fixed point. As we know from the Homogeneous
sine-Gordon models the $A_{n}|$\textbf{\~{g}}--theory may be viewed as
perturbed \textbf{\~{G}}$_{n+1}/U(1)^{\otimes \tilde{\ell}}$-coset WZNW
theories. In analogy, we could view for instance the ``dual'' theory of
this, i.e. the \textbf{\~{g}\TEXTsymbol{\vert}}$A_{n}$--theory, formally as
perturbed \textbf{\~{G}}$_{1}^{\otimes (n+1)}/$\textbf{\~{G}}$_{n+1}$-coset
WZNW theory. Besides the identification of the fixed point theory for the
situation in which \textbf{g}$\neq \!\!A_{n}$, it remains open to find the
precise form of the perturbing operators. We do not expect that they will
turn out to be irrelevant, since the colour giving CDD-factors are different
in nature than the ones recently discussed in \cite{Giu}. \bigskip

\textbf{Acknowledgments: } We are grateful to the Deutsche
Forschungsgemeinschaft (Sfb288) for financial support and B.J. Schulz,
O.A.~Castro-Alvaredo and J.L.~Miramontes for useful discussions.

\end{document}